\documentclass[sigconf]{acmart}
\DeclareUnicodeCharacter{FF0C}{,}

\AtBeginDocument{%
  \providecommand\BibTeX{{%
    \normalfont B\kern-0.5em{\scshape i\kern-0.25em b}\kern-0.8em\TeX}}}

\usepackage{booktabs}
\usepackage{multirow}
\usepackage{threeparttable}
\usepackage{comment}
\usepackage{subcaption}

\usepackage{xspace}
\usepackage{amsmath}
\usepackage{amsfonts}
\usepackage{caption}

\newcommand\approach{QuMAB}

\newcommand{\ie}{i.e.\@\xspace}
\newcommand{\eg}{e.g.\@\xspace}

\begin{document}

\title{QuMAB: Query-based Multi-Annotator Behavior Modeling with Reliability under Sparse Labels}

% \affiliation{%

% \author{Anonymous Authors}

% 设置作者信息
\author{Liyun Zhang}
\affiliation{%
  \institution{D3 Center, Osaka University}
  \country{Japan}
}
% \email{zhang.liyun@ids.osaka-u.ac.jp}

\author{Zheng Lian}
\affiliation{%
  \institution{Institute of automation, Chinese academy of science}
  \country{China}
}
% \email{lianzheng2016@ia.ac.cn}

\author{Hong Liu}
\affiliation{%
  \institution{Xiamen University}
  \country{China}
}
% \email{lynnliu.xmu@gmail.com}

\author{Takanori Takebe}
\affiliation{%
  \institution{Cincinnati Children's Hospital Medical Center}
  \country{Japan}
}
% \email{takanori.takebe@cchmc.org}

\author{Yuta Nakashima}
\affiliation{%
  \institution{D3 Center, Osaka University}
  \country{Japan}
}
% \email{n-yuta@ids.osaka-u.ac.jp}

%% You do not have to enter your paper ID


\begin{abstract}
Multi-annotator learning traditionally aggregates diverse annotations to approximate a single ``ground truth'', treating disagreements as noise. However, this paradigm faces fundamental challenges: subjective tasks often lack absolute ground truth, and sparse annotation coverage makes aggregation statistically unreliable. We introduce a paradigm shift from sample-wise aggregation to annotator-wise behavior modeling. By treating annotator disagreements as valuable information rather than noise, modeling annotator-specific behavior patterns can reconstruct unlabeled data to reduce annotation cost, enhance aggregation reliability, and explain annotator decision behavior.
To this end, we propose \approach\ (\textbf{\underline{Qu}}ery-based \textbf{\underline{M}}ulti-\textbf{\underline{A}}nnotator \textbf{\underline{B}}ehavior Pattern Learning), which uses lightweight queries to model individual annotators while capturing inter-annotator correlations as implicit regularization, preventing overfitting to sparse individual data while maintaining individualization and improving generalization, with a visualization of annotator focus regions offering an explainable analysis of behavior understanding. We contribute two large-scale datasets with dense per-annotator labels: STREET (4,300 labels/annotator) and AMER (average 3,118 labels/annotator), the first multimodal multi-annotator dataset.
Extensive experiments demonstrate the superiority of our \approach\ in modeling individual annotators' behavior patterns, their utility for consensus prediction, and applicability under sparse annotations.
\end{abstract}
%-------------------------------------------------------------------------



% \section{CCS Concepts and User-Defined Keywords}

% Two elements of the ``acmart'' document class provide powerful
% taxonomic tools for you to help readers find your work in an online
% search.

% The ACM Computing Classification System ---
% \url{https://www.acm.org/publications/class-2012} --- is a set of
% classifiers and concepts that describe the computing
% discipline. Authors can select entries from this classification
% system, via \url{https://dl.acm.org/ccs/ccs.cfm}, and generate the
% commands to be included in the \LaTeX\ source.

% User-defined keywords are a comma-separated list of words and phrases
% of the authors' choosing, providing a more flexible way of describing
% the research being presented.

% CCS concepts and user-defined keywords are required for for all
% articles over two pages in length, and are optional for one- and
% two-page articles (or abstracts).




% \subsection{The ``Teaser Figure''}
% A ``teaser figure'' is an image, or set of images in one figure, that
% are placed after all author and affiliation information, and before
% the body of the article, spanning the page. If you wish to have such a
% figure in your article, place the command immediately before the
% \verb|\maketitle| command:
% \begin{verbatim}
%   \begin{teaserfigure}
%     \includegraphics[width=\textwidth]{sampleteaser}
%     \caption{figure caption}
%     \Description{figure description}
%   \end{teaserfigure}
% \end{verbatim}

%   \centering
%   \includegraphics[width=\textwidth]{images/overview.jpg}
%   \caption{The application of thermal-to-color image translation }
%   \Description{Introducing our mission with a simulation diagram, so as to show it more vividly.}
%   \label{fig:introduction}

\maketitle

\section{Introduction}
\label{sec:intro}

\begin{figure}[t]
  \centering
  \includegraphics[width=1.0\linewidth]{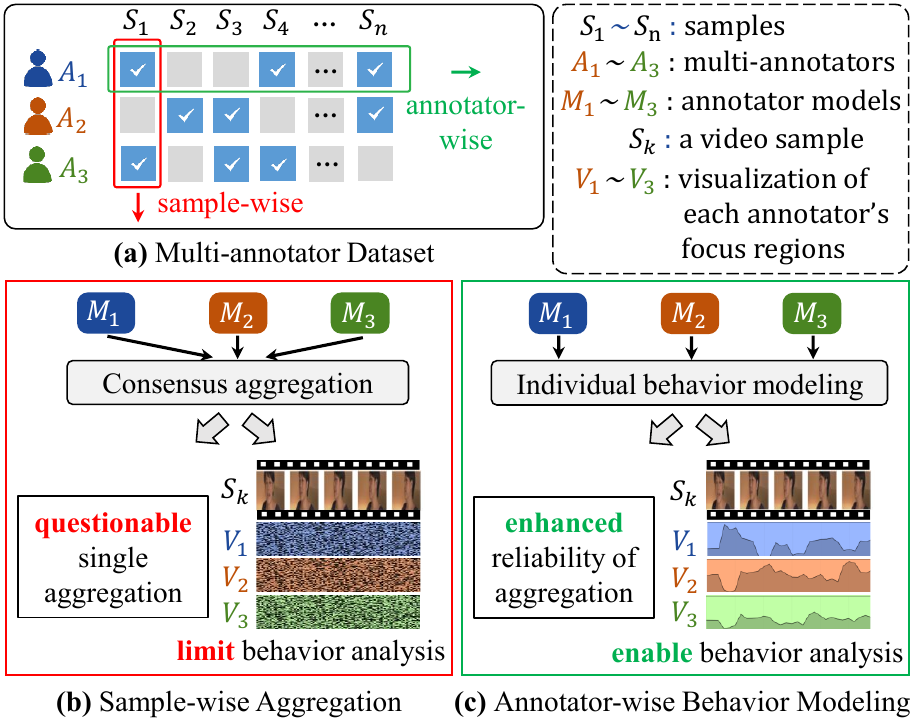}
  \caption{Paradigm shift from sample-wise aggregation to annotator-wise behavior modeling. \textbf{(a)}: Sparse annotation matrix showing each annotator labels a small subset of samples with disjoint coverage. \textbf{(b)}: Traditional sample-wise aggregation makes a questionable single ``ground truth'' prediction, potentially losing individual information to limit behavior analysis. \textbf{(c)}: Our annotator-wise behavior modeling captures each annotator's behavior patterns longitudinally across their labeled samples, enhancing reliability of aggregation via reconstructed unlabeled data in annotation matrix, offering explainable analysis of behavior understanding.}
  \label{overview}
\end{figure}

In real-world multi-annotation scenarios, such as medical image analysis \cite{PADL}, sentiment analysis \cite{lian2023mer}, and visual perception \cite{CNN-CM}, different annotators often provide different labels to the same sample \cite{LFC} due to different personal backgrounds, subjective interpretations, and preferences.
Traditional multi-annotator learning focuses on learning different characteristics (\eg, confusion mode \cite{Sampling-CM}, agreement \cite{Learn2agree}, expertise level \cite{Multi-rater}) from multiple annotators, then treating these discrepancies as bias or noise to elimant for achieving aggregation to approximate a single ``ground truth'' prediction \cite{noise1, Correction}.

However, the reliability of this paradigm faces two fundamental challenges: (1) In subjective domains such as emotional or impression assessment, there often exists no absolute ground truth—making this aggregation itself questionable \cite{LFC, CAF}. 
% There often exists no absolute ground truth to evaluate this prediction, particularly in subjective tasks such as emotional or impression assessment;
(2) In real-world crowdsourcing, each annotator labels only a small fraction of the data, with most samples receiving annotations from different, often disjoint annotator subsets. This sparse and fragmented coverage makes aggregation statistically unreliable, as there is insufficient overlap to establish robust consensus patterns \cite{Sparse}.
% As a result, consensus aggregation reflects only small, inconsistent subsets of annotators.

Therefore, we argue for a shift in focus, \ie, from sample-wise to annotator-wise (Figure~\ref{overview}): instead of treating sample-wise annotator disagreements as noise to be averaged away, we propose modeling individual annotator behavior patterns as annotator-wise valuable information. These patterns capture consistent differences in judgment arising from expertise, preference, or perspective.
By longitudinally learning reliable annotator-specific models—tracking each individual's behavior patterns across consecutive sample-label pairs per annotator rather than aggregating multiple annotators per sample—we unlock three compelling advantages: 
(1) \textbf{Cost reduction}: reconstructing unlabeled data enables comprehensive annotation coverage; 
% sufficient coverage of more annotators' sample-label pairs to significantly reduce the annotation cost;
(2) \textbf{Enhanced reliability}: aggregating over reconstructed sufficient coverage of sample-label pairs yields statistically more robust consensus than sparse, fragmented annotations;
% It enhance aggregation reliability through predicting consensus over reconstructed sufficient coverage of sample-label pairs rather than sparse;
(3) \textbf{Behavioral insights}: understanding the behavior patterns underlying each annotator's decisions and explaining sources of disagreement.
% revealing consistent patterns underlying each annotator's decisions explains why experts systematically disagree.
% It can help understanding the consistent patterns underlying each annotator's decisions and explaining persistent sources of disagreement. This is particularly valuable in high-stakes domains where understanding the source of disagreements is as important as the final decision.

% To this end, we seek an architecture that can well capture individual behavior patterns and enables a certain level of explainable analysis for behavior understanding in multi-annotator modeling.
Currently, existing research focusing on multi-annotator behavior pattern modeling and presenting explainable analysis of behavior understanding is sparse. Some works similar to it have attempted to model individual annotators through various techniques (\eg, MaDL's confusion matrices \cite{MaDL} or PADL's Gaussian distribution fitting \cite{PADL}) to understand more about individual annotator patterns. 
However, their aggregation-oriented mechanism (\eg, PADL's meta-learning, and MaDL's jointly optimizing consensus and annotator classifiers) may average annotator perspectives to lose annotator-specific information, influencing individual behavior modeling. Otherwise, if completely independent modeling individual annotator, it preserves sufficient individual information but easily suffers from overfitting during the annotator model is trained on its small set of labels as described previously. Moreover, existing models \cite{interpretable} lack explainable analysis of behavior understanding, or they only implicitly reveal trends where certain annotators play a larger role in the prediction \cite{TAX}.

Our key insight is that annotators, despite their individual differences, often share behavioral structures. By capturing inter-annotator correlations, we can leverage the collective patterns as implicit regularization, constraining individual models from overfitting while preserving unique characteristics.
% This constrains individual models from overfitting to their limited data while preserving each annotator's unique characteristics—achieving robust individual modeling even under sparse supervision.
To this end, we propose a novel query-based architecture \approach\ (\textbf{\underline{Qu}}ery-based \textbf{\underline{M}}ulti-\textbf{\underline{A}}nnotator \textbf{\underline{B}}ehavior Pattern Learning), hypothesize that annotator judgment differences arise from their varying degrees of focus on different regions of the input content (\eg, focusing on different image patches). Each annotator is represented by a learnable query that interacts with input features via cross-attention to effectively model individual behavior patterns. Lightweight query significantly reduces computational cost compared to separate conventional models.
Crucially, all annotator queries also interact through shared self-attention to capture inter-annotator correlations as a form of implicit structural regularization. This constrains inter-annotator representations to follow similarity patterns derived from annotations, preventing individual representations from drifting too far from the group and promoting mutual enhancement. This mechanism prevents overfitting to sparse individual data while maintaining individualization, improving generalization and robustness in individual annotator modeling, particularly under sparse annotations.
Additionally, the cross-attention weights provide a visualization of annotator focus regions, offering an explainable analysis of behavior understanding.

Furthermore, we contribute two new large-scale datasets with dense per-annotator labels: STREET (city impression assessment, 4,300 labels/annotator) and AMER (video emotion recognition, average 3,118 labels/annotator). These datasets provide a high-value longitudinal annotation perspective for understanding and evaluating individual annotator behavior patterns, offering valuable data to the community and further researchers. It is worth noting that AMER is the first multi-annotator multimodal dataset in this field. Our work makes the following contributions:

\begin{itemize}
  \item \textbf{A paradigm focus shift in multi-annotator learning}: We introduce a paradigm shift from sample-wise consensus aggregation to annotator-wise behavior modeling. By treating annotator disagreements as valuable information rather than noise, modeling annotator-specific behavior patterns can reconstruct unlabeled data to reduce annotation cost, enhance aggregation reliability, and explain annotator behavior.
  
  \item \textbf{A novel query-based architecture}: We propose \approach, which uses lightweight queries to model individual annotators while capturing inter-annotator correlations as implicit regularization, preventing overfitting to sparse individual data while maintaining individualization and improving generalization, with a visualization of annotator focus regions offering an explainable analysis of behavior understanding.

  \item \textbf{Two new large-scale datasets}: We contribute STREET (4,300 labels per annotator) and AMER (average 3,118 labels per annotator) datasets with denser per-annotator labels than existing resources, offering a longitudinal perspective for understanding individual annotator behaviors. AMER is the first multimodal multi-annotator dataset.
\end{itemize}
%-------------------------------------------------------------------------

\begin{table*}[htbp]
    \centering
    % \small
    \caption{Dataset comparison. Compared to existing datasets, our datasets contain a greater number of samples annotated by each annotator, helping promote multi-annotator behavior pattern modeling. AMER is the first multimodal multi-annotator dataset.}
    \label{tab:dataset_comparison}
    \begin{tabular}{llcc}
    \toprule
    Dataset & Dataset description  & Modality & \# samples per annotator  \\
    \toprule
    QUBIQ-kidney \cite{menze2020quantification} & kidney image & image & 24 \\
    QUBIQ-tumor \cite{menze2020quantification} & brain tumor image & image & 32\\
    QUBIQ-growth \cite{menze2020quantification} & brain growth image & image & 39\\
    QUBIQ-prostate \cite{menze2020quantification} & prostate image  & image & 55\\
    CIFAR-10H \cite{peterson2019human} & object recognition  & image & 200 \\
    MUSIC \cite{rodrigues2013learning} & music genre classification  & audio & 2$\sim$368\\
    MURA \cite{rajpurkar2017mura} & radiographic image & image & 556 \\
    RIGA \cite{almazroa2017agreement} & retinal cup and disc segmentation & image & 750\\
    LIDC-IDRI \cite{armato2011lung} & lung nodule image  & image & 1,018\\
    %GHC \cite{kennedy2018gab} & hate speech classification & audio  & 288$\sim$13,543 \\
    \midrule
    \textbf{STREET (Ours)} & city impression evaluation & image  & 4,300 \\
    \textbf{AMER (Ours)} & video emotion recognition & audio, video, text  & 970$\sim$5,202\\
    \bottomrule
    \end{tabular}
\end{table*}

\section{Related Work}

\subsection{Multi-annotator Behavior Modeling Paradigm}
To the best of our knowledge, the multi-annotator behavior modeling paradigm problem has not yet been investigated. Traditional multi-annotator learning focuses on estimating consensus or ground-truth labels from multiple noisy annotations. These include early probabilistic models \cite{DS_model}, EM algorithms \cite{GLAD}, Gaussian models \cite{GP-MLL}, and biased estimation \cite{bias_annotator}. Tanno et al. \cite{tanno2019learning} proposed modeling annotator confusion matrices as learnable parameters in neural networks. Cao et al. \cite{cao2019max} introduced max-MIG to learn from multiple annotators. SimLabel~\cite{SimLabel} addresses the practical challenge of missing labels in multi-annotator scenarios. NEAL \cite{NEAL} employs neural expectation-maximization to jointly learn annotator expertise and true labels. Later methods used probabilistic frameworks to aggregate multiple annotations into a consensus or ground-truth label by confusion matrix \cite{Sampling-CM}, agreement distribution \cite{Learn2agree}, and Gaussian distributions \cite{PADL}. This sample-wise aggregation paradigm often treats annotator disagreements as noise to be averaged away rather than valuable information \cite{noise1, Correction}.
In contrast, our introduced annotator-wise modeling paradigm treats annotator disagreements as valuable information for modeling annotator-specific behavior patterns, enhancing aggregation reliability, and explaining annotator behavior.

\subsection{Multi-annotator Behavior Modeling Architecture}
Previous studies have attempted to model individual annotators through various techniques: D-LEMA \cite{D-LEMA} trains annotator models on non-contradictory subsets with spatial weights for noise handling; PADL \cite{PADL} models annotator preferences via Gaussian assumptions in its Human Preference Module (HPM) and employs a Sample Embedding Module (SEM) for meta-classification; MaDL \cite{MaDL} jointly optimizes ground truth classifiers and annotator models through weighted embeddings. However, their aggregation-oriented mechanisms compromise individual behavior modeling by averaging annotator perspectives: D-LEMA sacrifices annotator-specific patterns for fusion objectives, PADL forces individual distributions to converge at the cost of behavioral details, and MaDL smooths individual characteristics to minimize consensus loss. 
Meanwhile, existing efforts on explainable analysis of annotator behavior understanding remain limited. Some works provide insights, \eg, TAX \cite{TAX} associates convolutional kernels with prototype libraries for pixel-level annotation decisions, MAGI \cite{MAGI} leverages annotator explanations to address noisy annotations, and Schaekermann et al.~\cite{structured_adjudication} analyze factors contributing to disagreements. However, they only reveal annotators' trends in aggregation or analyze isolated factors without behavioral analysis. 
In comparison, our architecture models individual annotators via lightweight queries, leveraging inter-annotator correlations as regularization against overfitting while preserving individualization, with attention visualization analyzing behavioral patterns.

\subsection{Multi-annotator Datasets}
Most existing multi-annotator datasets often have only a small subset of samples with consecutive annotations from consistent annotator IDs. CIFAR-10H \cite{CIFAR-10H}, based on the CIFAR-10 \cite{CIFAR-10} dataset, includes 10,000 test samples labeled by 2,571 annotators, but each annotator ID has on average only about 200 consecutive labels. LabelMe \cite{Labelme} includes an average of approximately 42.4 consecutive labels per annotator ID. Audio dataset Music \cite{Music} contains an average of about 46.1 consecutive labels. The medical datasets are commonly used in multi-annotator studies, and consecutive annotator labels are even sparser. QUBIQ \cite{QUBIQ}, a dataset for quantifying uncertainty in biomedical image segmentation, includes four distinct segmentation datasets with an average of only 40 samples, and even then, annotator IDs have only around 8 consecutive labeled samples each. For longitudinal annotator behavior understanding, we contribute two new large-scale datasets with dense per-annotator labels: STREET (city impression assessment, 4,300 labels/annotator) and AMER (video emotion recognition, average 3,118 labels/annotator). AMER is the first multimodal multi-annotator dataset.
%-------------------------------------------------------------------------

\begin{figure*}[t]
  \centering
  \includegraphics[width=0.8\linewidth]{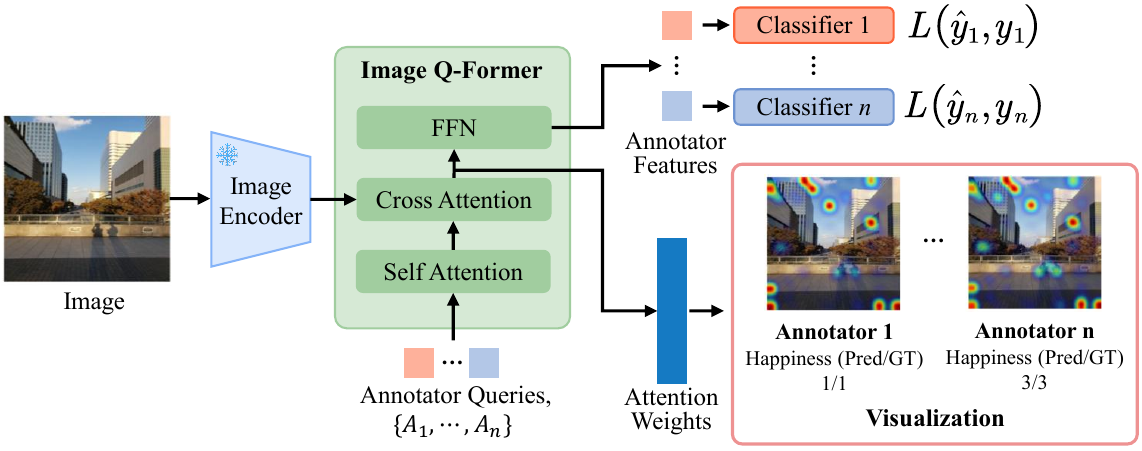}
    \caption{\approach\ for image. A frozen pre-trained image encoder extracts features, which interact with annotator-specific learnable queries in the Image Q-Former through cross-attention, producing annotator-specific features for classification. Different annotators’ cross-attention weights reflect the differences in the image patches they focus on and support visual interpretation.}
    % \approach\ architecture for video-specific pipeline is provided in the supplementary material.
    \label{fig:architecture-image}
\end{figure*}

\section{Dataset Construction}
\label{sec:dataset}
We contribute two new large-scale datasets\footnote{We are negotiating to publish both datasets after acceptance.}: STREET (city impression assessment) and AMER (multi-modal emotion recognition) in this paper. Table \ref{tab:dataset_comparison} compares the current multi-annotator datasets. We observe that in existing datasets, the number of samples annotated by each annotator is relatively small, and there is a lack of multi-annotator multimodal datasets. For example, in the RIGA dataset \cite{almazroa2017agreement}, each annotator labels 750 samples, while in the CIFAR-10H dataset \cite{peterson2019human}, each annotator labels 200 samples.

{\textbf{(1) STREET}} is an urban perception dataset with multi-annotators, which contains $4,300$ high-resolution images covering various urban elements, such as streets, public spaces, and infrastructure. The images were captured during a series of city strolling surveys, which aim to analyze emotions in relation to various factors associated with the city. The surveys were conducted by an organization to which one of our co-authors belongs. Voluntary participants walked around their own familiar city, took photos of various factors that may affect their subjective feelings (i.e., happiness/health), and assigned labels related to these feelings to each image (though we do not use these labels but ones assigned by crowd workers in our experiments). Thirteen survey sessions were conducted in five different cities (three urban areas and two suburban areas). A total of 327 participants, ranging in age from their 10s to 60s, took part in the survey. Each session lasted about one hour, with each participant taking an average of 12.6 photos. We outsourced the annotation process to a company, which selected $10$ annotators with balanced age, gender, and location diversity on a platform similar to Amazon Mechanical Turk, assessing five perception dimensions: happiness, healthiness, safety, liveliness, and orderliness, using a $6$-point scale ($-3$ to $+3$). Each annotator spent approximately three weeks on their annotations. This multi-annotator dataset provides comprehensive human perception data for urban environments, enabling the quantitative analysis of environmental features and their emotional impact.

{\textbf{(2) AMER}} is a multimodal emotion dataset; the raw data is sourced from MER2024 \cite{lian2024mer}, which contains $5,207$ video samples from movies and TV series, with multi-annotator emotion labels. Each sample typically contains one person, with relatively complete speech content. We annotated AMER using the open-source software Label Studio \cite{LabelStudio}. We hired $15$ annotators, who were students of our co-author's institution, and underwent a training session with $10$ samples. We retained $13$ annotators after screening out careless and irresponsible ones. Each annotator completed the task in approximately two weeks, with scheduled breaks to maintain annotation quality, where each annotator selects the most likely label from 8 candidate labels, i.e., worry, happiness, neutrality, anger, surprise, sadness, other, and unknown. Among all annotators, $10$ annotators show consistent participation, each providing approximately $970$ to $1,096$ labels, while the remaining 3 annotators contribute over $5,000$ labels each. This rich multi-annotator setup provides reliable emotion annotation results and allows for a robust evaluation of emotion recognition performance.
%-------------------------------------------------------------------------  

\section{Methodology}
\label{sec:method}
We propose \approach, a query-based architecture designed to model the behavior patterns of individual annotators.
We take the image classification task to illustrate our image-specific architecture for image inputs from the STREET dataset, which consists of a frozen image encoder, annotator-specific learnable queries, an image Q-Former, and annotator-wise classifiers, as shown in Figure~\ref{fig:architecture-image}. For video inputs from the AMER dataset, we describe a video-specific architecture in the supplementary material.

Given an image input $I \in \mathbb{R}^{H \times W \times 3}$, a frozen pre-trained image encoder~\cite{EVA-CLIP} first extracts image features, which are then fed into the Image Q-Former~\cite{BLIP2}. For each annotator $A_k$ ($k = 1, \dots, n$), we assign a learnable query token for modeling. These queries are first processed by a shared self-attention layer to allow them interact with each other to implicitly capture inter-annotator correlations, and then interact with image features through multi-head cross-attention (typically with 12 heads), enabling each query to access diverse attention perspectives and produce individualized representations that reflect each annotator's potential focus and decision process. The resulting representations are mapped through a fully connected layer and passed to each annotator's classifier for prediction.

This query-based design is motivated by the hypothesis that annotator judgment differences arise from their varying degrees of focus on different regions of the input content (\eg, focusing on different image patches). Through learnable queries and cross-attention, our model effectively captures these individualized behavior patterns.
Additionally, representing each annotator with a lightweight query significantly reduces computational cost compared to separate conventional models.

Meanwhile,  the mechanism of capturing inter-annotator correlations acts as a form of implicit structural regularization. This constrains inter-annotator representations to follow similarity patterns derived from annotations, preventing individual representations from drifting too far from the group and promoting mutual enhancement. It also prevents overfitting to annotator-specific noise while preserving individual differences of behavior patterns. As a result, the model achieves improved generalization and robustness in individual annotator modeling, particularly under sparse annotations.

For qualitative understanding, we visualize the cross-attention weights from the Image Q-Former to reveal annotator-specific focus regions. These weights indicate which image patches different annotators focus on when making predictions.
Figure~\ref{fig:architecture-image} illustrates results on the STREET dataset (city impression classification), annotators exhibit distinct spatial focus: \emph{annotator n} attends more strongly to the two people holding hands in the center. This contrast reflects how annotators may interpret emotional cues differently based on their focus: \emph{annotator n} assigns a higher score to the ``happiness'' dimension compared to \emph{annotator 1}, suggesting that variations in focus on semantically positive regions may contribute to their differing judgments.

\subsection{Loss Function}
Finally, as shown in Figure~\ref{fig:architecture-image}, the total training loss $\mathcal{L}_{\text{total}}$ for the proposed multi-annotator classification model, \approach, is defined as the sum of individual cross-entropy losses for each annotator:
\begin{equation}
\mathcal{L}_{\text{total}} = \sum_{k=1}^{n} \mathcal{L}(\hat{y}_{k}, y_{k}),
\end{equation}
where each annotator $A_{k}$ has a specific predicted probabilities $\hat{y}_{k} \in [0, 1]^C$, a reference label $y_{k} \in \{0, 1\}^C$ in one-hot vector representation, and $C$ is the number of classes.
%-------------------------------------------------------------------------

\section{Experiment}
\label{sec:experiment}
We conduct extensive experiments to evaluate our \approach, including modeling individual annotators' behavior patterns, assessing their utility for consensus prediction, testing applicability under sparse annotations, and complementing with qualitative visualization analysis. We compare against three representative baselines: D-LEMA~\cite{D-LEMA}, an ensemble-based multi-annotator learner; PADL~\cite{PADL}, which fits Gaussian distributions for each annotator; and MaDL~\cite{MaDL}, which models annotator-specific confusion matrices. To ensure our model captures annotator-specific patterns rather than shared encoder features, we also include a Base variant with only the encoder and classifiers. We evaluate on two multi-annotator datasets: AMER (video-based emotion recognition with 13 annotators) and STREET (urban image impression assessment with 10 annotators across five dimensions: Happiness, Healthiness, Safety, Liveliness, and Orderliness). The AMER dataset's complexity in capturing temporal emotion dynamics aligns with recent advances in time-sensitive emotion recognition~\cite{MicroEmo-arxiv, MicroEmo-mm}. These datasets provide dense, diverse annotations crucial for modeling annotator-specific behavior patterns. Accuracy and $F_1$ score \cite{F1-score} are used as evaluation metrics. Note that additional experiments, including model efficiency, a faithfulness-oriented interpretability analysis, extended results, and further discussion, are provided in the supplementary material.

\begin{table*}
\centering
% \small
\caption{The accuracy (ACC) and $F_1$ score evaluate results on the AMER dataset. We assess performance for individual annotator modeling (each annotator $A_{k}$, \textit{k} = 1, \dots, 13), the average (Avg), and consensus prediction (CoPr). Higher is better.}
\label{tab:amer_accuracy_f1}
\begin{tabular}{clccccccccccccccc}
\toprule
Metric & Methods & $A_1$ & $A_2$ & $A_3$ & $A_4$ & $A_5$ & $A_6$ & $A_7$ & $A_8$ & $A_9$ & $A_{10}$ & $A_{11}$ & $A_{12}$ & $A_{13}$ & Avg & CoPr \\
\midrule
\multirow{5}{*}{ACC} 
& Base & 0.30 & 0.29 & 0.43 & 0.25 & 0.24 & 0.20 & 0.26 & 0.19 & 0.34 & 0.10 & 0.41 & 0.48 & 0.19 & 0.28 & 0.35 \\
& D-LEMA & 0.86 & 0.88 & 0.85 & 0.87 & 0.89 & 0.86 & 0.88 & 0.87 & 0.85 & 0.86 & 0.45 & 0.51 & 0.33 & 0.78 & 0.55 \\
& PADL & 0.89 & 0.90 & 0.88 & 0.93 & 0.87 & 0.91 & 0.86 & 0.94 & 0.89 & 0.88 & 0.47 & 0.54 & 0.35 & 0.79 & 0.52 \\
& MaDL & 0.93 & 0.91 & 0.90 & 0.89 & 0.90 & 0.88 & 0.90 & 0.89 & 0.87 & 0.92 & 0.50 & 0.53 & 0.37 & 0.80 & 0.57 \\
& Ours & \textbf{0.94} & \textbf{0.93} & \textbf{0.93} & \textbf{0.94} & \textbf{0.94} & \textbf{0.92} & \textbf{0.93} & \textbf{0.95} & \textbf{0.93} & \textbf{0.93} & \textbf{0.59} & \textbf{0.61} & \textbf{0.40} & \textbf{0.84} & \textbf{0.60} \\
\midrule
\multirow{5}{*}{$F_1$} 
& Base & 0.26 & 0.27 & 0.40 & 0.22 & 0.23 & 0.17 & 0.24 & 0.18 & 0.31 & 0.08 & 0.35 & 0.41 & 0.14 & 0.25 & 0.32 \\
& D-LEMA & 0.84 & 0.87 & 0.81 & 0.84 & 0.86 & 0.85 & 0.86 & 0.82 & 0.83 & 0.84 & 0.38 & 0.44 & 0.27 & 0.73 & 0.52 \\
& PADL & 0.86 & 0.88 & 0.85 & 0.91 & 0.83 & 0.89 & 0.82 & 0.92 & 0.85 & 0.86 & 0.41 & 0.50 & 0.29 & 0.76 & 0.49 \\
& MaDL & 0.90 & 0.85 & 0.87 & 0.86 & 0.87 & 0.85 & 0.87 & 0.86 & 0.84 & 0.92 & 0.45 & 0.48 & 0.33 & 0.77 & 0.54 \\
& Ours & \textbf{0.91} & \textbf{0.91} & \textbf{0.90} & \textbf{0.92} & \textbf{0.91} & \textbf{0.90} & \textbf{0.89} & \textbf{0.93} & \textbf{0.91} & \textbf{0.93} & \textbf{0.54} & \textbf{0.55} & \textbf{0.34} & \textbf{0.81} & \textbf{0.57} \\
\bottomrule
\end{tabular}
\end{table*}

\begin{table*}
\centering
% \small
\caption{The accuracy metric is to evaluate results on the STREET dataset. We assess performance for individual annotator modeling (each annotator $A_{k}$, \textit{k} = 1, \dots, 13), the average (Avg), and consensus prediction (CoPr). Higher is better.}
\label{tab:street_accuracy}
\begin{tabular}{llcccccccccccc}
\toprule
Perspectives & Methods & $A_1$ & $A_2$ & $A_3$ & $A_4$ & $A_5$ & $A_6$ & $A_7$ & $A_8$ & $A_9$ & $A_{10}$ & Avg & CoPr \\
\midrule
\multirow{4}{*}{Happiness} & Base & 0.80 & 0.12 & 0.27 & 0.38 & 0.56 & 0.35 & 0.44 & 0.44 & 0.31 & 0.55 & 0.42 & 0.45 \\
& D-LEMA & 0.85 & 0.71 & 0.44 & 0.36 & \textbf{0.70} & 0.43 & 0.46 & 0.54 & 0.41 & 0.47 & 0.54 & 0.57 \\
 & PADL & 0.93 & 0.74 & 0.48 & 0.53 & 0.57 & 0.47 & 0.42 & 0.51 & 0.50 & 0.60 & 0.58 & 0.55 \\
 & MaDL & 0.91 & 0.77 & 0.44 & 0.38 & \textbf{0.70} & 0.47 & 0.46 & \textbf{0.54} & 0.48 & 0.47 & 0.56 & 0.58 \\
 & Ours & \textbf{0.94} & \textbf{0.80} & \textbf{0.54} & \textbf{0.55} & 0.69 & \textbf{0.51} & \textbf{0.53} & \textbf{0.54} & \textbf{0.52} & \textbf{0.64} & \textbf{0.63} & \textbf{0.62} \\
\midrule
\multirow{4}{*}{Healthiness} & Base & 0.77 & 0.19 & 0.14 & 0.47 & 0.87 & 0.36 & 0.43 & 0.44 & 0.51 & 0.54 & 0.47 & 0.56 \\
 & D-LEMA & 0.83 & \textbf{0.77} & 0.44 & 0.43 & 0.87 & 0.41 & 0.44 & 0.46 & 0.55 & 0.46 & 0.57 & 0.54 \\
 & PADL & \textbf{0.92} & 0.72 & 0.55 & 0.44 & 0.84 & 0.47 & 0.46 & 0.44 & 0.52 & 0.55 & 0.59 & 0.50 \\
 & MaDL & 0.89 & \textbf{0.77} & 0.44 & 0.44 & \textbf{0.90} & 0.41 & 0.44 & 0.46 & 0.55 & 0.46 & 0.58 & 0.55 \\
 & Ours & \textbf{0.92} & 0.75 & \textbf{0.56} & \textbf{0.52} & \textbf{0.90} & \textbf{0.49} & \textbf{0.48} & \textbf{0.54} & \textbf{0.64} & \textbf{0.61} & \textbf{0.64} & \textbf{0.58} \\
\midrule
\multirow{4}{*}{Safety} & Base & 0.58 & 0.65 & 0.36 & 0.36 & 0.61 & 0.37 & 0.56 & 0.40 & 0.32 & 0.53 & 0.47 & 0.51 \\
 & D-LEMA & 0.62 & 0.69 & 0.27 & 0.41 & 0.50 & 0.46 & 0.48 & 0.40 & 0.35 & 0.50 & 0.47 & 0.49 \\
 & PADL & \textbf{0.72} & 0.78 & 0.24 & 0.44 & 0.69 & 0.44 & \textbf{0.53} & 0.42 & 0.46 & 0.48 & 0.52 & 0.54 \\
 & MaDL & 0.63 & 0.63 & 0.27 & 0.32 & 0.61 & 0.38 & 0.46 & 0.42 & 0.36 & 0.52 & 0.46 & 0.56 \\
 & Ours & \textbf{0.72} & \textbf{0.80} & \textbf{0.38} & \textbf{0.48} & \textbf{0.71} & \textbf{0.54} & \textbf{0.53} & \textbf{0.50} & \textbf{0.52} & \textbf{0.58} & \textbf{0.58} & \textbf{0.61} \\
\midrule
\multirow{4}{*}{Liveliness} & Base & 0.79 & 0.60 & 0.53 & 0.46 & 0.76 & 0.30 & 0.35 & 0.36 & 0.53 & 0.57 & 0.53 & 0.55 \\
 & D-LEMA & 0.79 & 0.58 & 0.37 & 0.42 & 0.74 & 0.38 & 0.44 & 0.41 & 0.50 & 0.46 & 0.51 & 0.53 \\
 & PADL & 0.85 & 0.66 & 0.56 & 0.46 & 0.75 & 0.44 & 0.43 & 0.47 & 0.56 & 0.57 & 0.58 & 0.54 \\
 & MaDL & 0.78 & 0.56 & 0.35 & 0.40 & 0.76 & 0.34 & \textbf{0.48} & 0.42 & 0.47 & 0.47 & 0.50 & 0.56 \\
 & Ours & \textbf{0.87} & \textbf{0.68} & \textbf{0.57} & \textbf{0.53} & \textbf{0.80} & \textbf{0.49} & \textbf{0.48} & \textbf{0.51} & \textbf{0.62} & \textbf{0.61} & \textbf{0.62} & \textbf{0.59} \\
\midrule
\multirow{4}{*}{Orderliness} & Base & 0.50 & 0.64 & 0.39 & 0.45 & 0.86 & 0.31 & 0.39 & 0.34 & 0.31 & 0.49 & 0.47 & 0.52 \\
 & D-LEMA & 0.55 & 0.60 & 0.32 & 0.36 & 0.82 & 0.39 & 0.42 & 0.36 & 0.37 & 0.47 & 0.47 & 0.57 \\
 & PADL & 0.73 & 0.65 & 0.44 & 0.45 & 0.93 & 0.45 & 0.45 & 0.36 & 0.42 & \textbf{0.63} & 0.55 & 0.54 \\
 & MaDL & 0.61 & 0.60 & 0.34 & 0.36 & 0.86 & 0.37 & 0.47 & 0.36 & 0.37 & 0.49 & 0.48 & 0.58 \\
 & Ours & \textbf{0.74} & \textbf{0.71} & \textbf{0.52} & \textbf{0.55} & \textbf{0.94} & \textbf{0.47} & \textbf{0.54} & \textbf{0.44} & \textbf{0.56} & 0.62 & \textbf{0.61} & \textbf{0.62} \\
\bottomrule
\end{tabular}
\end{table*}

\begin{table}[htbp]
\centering
% \small
\caption{Evaluation by accuracy for sparse scenarios (40\% of annotations are randomly removed). S-Ha, S-He, S-Sa, S-Li, and S-Or represent five perspectives of STREET dataset: happiness, healthiness, safety, liveliness, and orderliness.}
\label{tab:sparse}
\begin{tabular}{l@{\hspace{2.8mm}}ccccc@{\hspace{3mm}}c}
\toprule
Method & S-Ha & S-He & S-Sa & S-Li & S-Or & AMER \\
\midrule
Full-PADL & 0.58 & 0.59 & 0.52 & 0.58 & 0.55 & 0.79 \\
Full-Ours & 0.63 & 0.64 & 0.58 & 0.62 & 0.61 & 0.84 \\
% \midrule
% 20\%-PADL & 0.54 & 0.54 & 0.48 & 0.53 & 0.50 & 0.73 \\
% 20\%-Ours & 0.60 & 0.60 & 0.54 & 0.58 & 0.56 & 0.79 \\
% \midrule
% 30\%-PADL & 0.50 & 0.49 & 0.44 & 0.48 & 0.45 & 0.67 \\
% 30\%-Ours & 0.57 & 0.56 & 0.51 & 0.54 & 0.52 & 0.73 \\
\midrule
Sparse-PADL & 0.43 & 0.42 & 0.38 & 0.41 & 0.40 & 0.58 \\
Sparse-Ours & \textbf{0.52} & \textbf{0.51} & \textbf{0.46} & \textbf{0.49} & \textbf{0.48} & \textbf{0.66} \\
\bottomrule
\end{tabular}
\end{table}

\subsection{Implementation Details}
Our image-specific model pipeline (Figure~\ref{fig:architecture-image}) uses ViT-G/14 from EVA-CLIP \cite{EVA-CLIP} as the encoder, with Image Q-Former initialized from InstructBLIP \cite{InstructBLIP} (Frame Q-Former is the same in a video-specific pipeline from supplementary material, where Video Q-Former is initialized from Video-LLaMA \cite{Video-llama}).
Input images and video frames are resized to 224$\times$224 and normalized. The number of query tokens is set equal to the number of annotators, and each annotator's classifier model uses an MLP.
We train the model using the AdamW optimizer with an initial learning rate of 1e-4, a weight decay of 0.01, and gradient clipping with a maximum norm of 1.0. A linear warmup strategy is applied for the first 20\% steps followed by cosine learning rate decay. The model is trained for up to 200 epochs with early stopping (patience = 25) to avoid overfitting.
Training is conducted using distributed data parallelism (DDP) on four NVIDIA V100 GPUs.

\begin{figure*}
  \centering
  \includegraphics[width=0.85\linewidth]{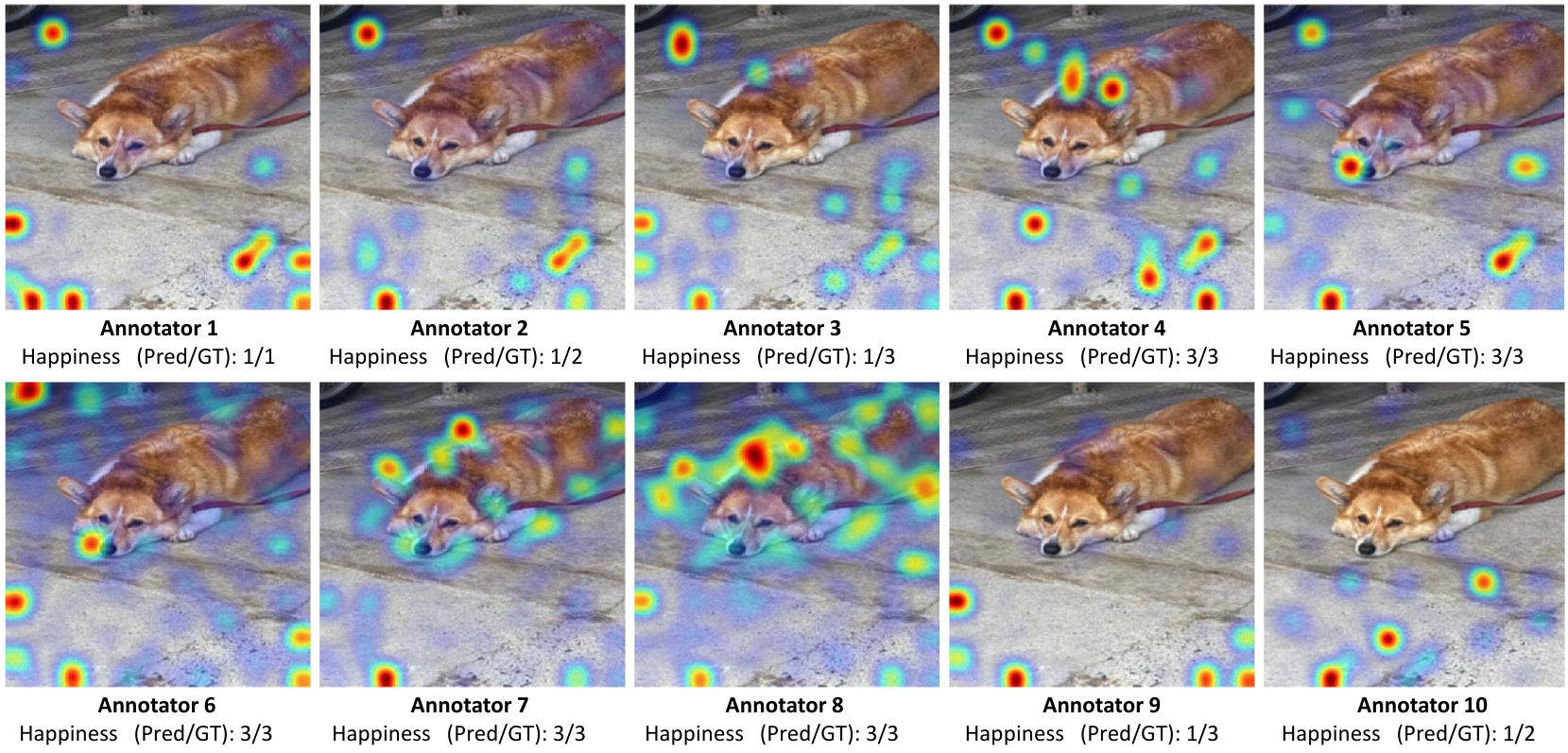}
  \caption{A visualization analysis of the different image patches that annotators focused on in the STREET dataset. The \emph{annotators 4, 5, 6, 7, and 8} exhibit centralized focuses on a cute dog compared to other annotators.}
  \label{fig:S1}
\end{figure*}

\begin{figure*}
  \centering
  \includegraphics[width=0.85\linewidth]{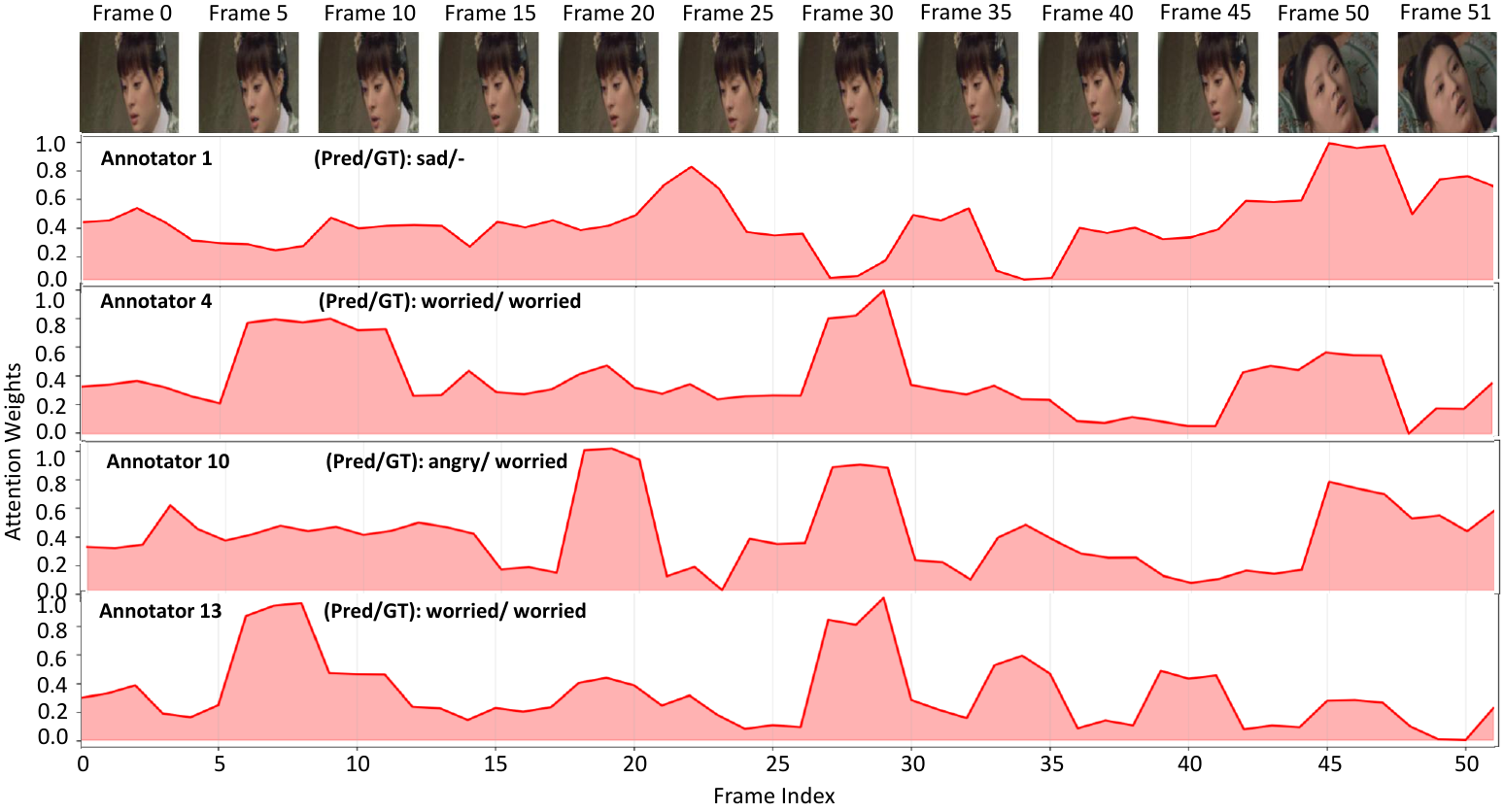}
  \caption{A visualization analysis of the different video frames that annotators focused on in the AMER dataset. The \emph{annotator 1} exhibits focus on final frames (the frames contain different person), while \emph{annotators 4 and 13} focus on the middle frames.}
  \label{fig:A1}
\end{figure*}

% \begin{figure*}
%   \centering
%   \begin{subfigure}{1.0\linewidth}
%     \centering
%     \includegraphics[width=0.9\linewidth]{images/file7.pdf}
%     \caption{An image from the STREET dataset.}
%     \label{fig:S1}
%   \end{subfigure}

%   \vspace{1em}

%   \begin{subfigure}{1.0\linewidth}
%     \centering
%     \includegraphics[width=0.9\linewidth]{images/file3.pdf}
%     \caption{A video from the AMER dataset.}
%     \label{fig:A1}
%   \end{subfigure}
   
  % \caption{We provide the visual interpretation of the differences in the image patches \protect\subref{fig:S1} and video frames \protect\subref{fig:A1} that different annotators focus on, reflected by their different cross-attention weights. In \protect\subref{fig:S1}, \emph{annotators 4, 5, 6, 7, and 8} exhibit centralized focuses on a cute dog compared to other annotators. In \protect\subref{fig:A1}, the \emph{annotator 1} exhibits focus on final frames (the frames contain different person), while \emph{annotators 4 and 13} focus on the middle frames.}
%   \label{fig:va}
% \end{figure*}

\subsection{Evaluation Metrics}
To evaluate the performance of individual annotator modeling and consensus prediction (majority-votes the predictions of multi-annotators), we use accuracy (a standard metric in the multi-annotator learning) and $F_1$ score \cite{F1-score} balancing precision and recall, suitable for potential class imbalance from uneven annotation densities, as in AMER (1,040 vs. 5,195 labels for annotators 1–10 vs. 11–13).
% For the performance evaluation of individual annotators' behavior patterns modeling and consensus prediction, we use accuracy (standard metric in the multi-annotator learning) and $F_1$ score \cite{F1-score} balancing precision and recall, suitable for potential class imbalance from uneven annotation densities, as in AMER (1,040 vs. 5,195 labels for annotators 1–10 vs. 11–13).
% For model interpretability, we use the comprehensiveness score~\cite{comprehensiveness, comparative—faithfulness} to quantitatively assess the faithfulness of our attention-based explanations.

\subsection{Quantitative Results}
We analyze evaluation results on modeling individual annotators' behavior patterns, their utility for consensus application, and applicability under sparse annotation scenarios.

\textbf{Individual Annotator Modeling.}
Individual annotator modeling aims to capture the behavior patterns of different annotators. Results in Tables~\ref{tab:amer_accuracy_f1} and \ref{tab:street_accuracy} show that our method consistently outperforms all baselines in both accuracy and $F_1$ score \footnote{\label{street-f1}See supplementary material for $F_1$ results on STREET dataset.} across individual annotators on the STREET and AMER datasets.
%(Detailed Table results of $F_1$ scores on the STREET dataset are provided in supplementary material due to space limitation, where we achieve overall average higher scores, \eg, 0.61 on the healthiness perspective compared to 0.56 for he best baseline PADL) 
%, although AMER has varying annotation densities across annotators.
This validates the superiority of our approach in capturing individual annotator behavior patterns.

\textbf{Consensus Application Benefits.}
Real-world applications often seek a single consensus label despite subjectivity among annotators. We evaluate consensus prediction to validate whether modeling individual annotators preserves valuable information for practical needs. As no definitive ground truth exists, we use majority vote over raw annotations as a proxy, acknowledging its potential biases.
Existing baselines adopt different aggregation strategies: D-LEMA learns weighted fusion; PADL applies meta-learning; and MaDL jointly optimizes consensus and annotator classifiers. They may average annotator perspectives during training, potentially diminishing individual nuances.
For a fair comparison, we apply unified majority voting over annotator-specific predictions from all methods rather than using their original aggregated outputs. Results (CoPr) in Tables~\ref{tab:amer_accuracy_f1}, \ref{tab:street_accuracy}, and $F_1$ results \footref{street-f1} show that our method achieves superior consensus performance, suggesting that modeling individual annotators helps retain valuable information, potentially benefiting real-world consensus applications.
% , even without explicit consensus optimization.
\textit{Note:} This experiment serves to validate practical utility of individual annotator modeling rather than to assert overall superiority.
% \textbf{Consensus Prediction.}  
% Consensus prediction aims to produce a single reliable label despite subjective annotations. Since no ground truth exists, we evaluate all methods using majority vote as proxy ground truth. Existing baselines employ different aggregation strategies: Base uses simple majority voting; D-LEMA learns weighted ensemble fusion; PADL employs meta-learning for consensus; MaDL jointly optimizes consensus and annotator classifiers. However, these methods average annotator perspectives during joint training—D-LEMA through weighted fusion objectives, PADL through meta-classification constraints, and MaDL through shared optimization goals—potentially losing individual annotator nuances. For fair comparison, we discard their original aggregation modules and apply unified majority voting over each method's annotator-specific predictions. Results in Tables~\ref{tab:amer_accuracy_f1} and \ref{tab:street_accuracy} show our method achieves superior consensus performance (e.g., 0.60 accuracy, 0.57 $F_1$ on AMER vs. 0.57/0.54 for MaDL), demonstrating that preserving individual annotator modeling enhances consensus quality. This validates our hypothesis that explicit annotator-specific learning without joint aggregation constraints better captures diverse perspectives for reliable consensus prediction.
% \textbf{Important Note:} This experiment is not intended to compare method superiority, but rather to validate the general principle that individual annotator modeling enhances information retention across different architectural approaches.

\textbf{Applicability under Sparse Annotations.}
To evaluate our model’s applicability under sparse annotation scenarios, we simulated real-world conditions by randomly removing annotations at various rates. As shown in Table~\ref{tab:sparse}, when 40\% of annotations are removed\footnote{\label{sparse-rate}See supplementary material for results of more sparse rates.}, our model’s average performance drops by 20.4\%, whereas the best baseline PADL experiences a larger drop of 27.4\%.
Results suggest that our superiority stems from modeling inter-annotator correlations, which regularizes individual annotator representations, preventing overfitting to sparse labels and promoting consistency with shared patterns across annotators, to enhance robustness and generalization under sparse annotations.
% These results suggest that modeling inter-annotator correlations and leveraging shared supervisory signals help maintain performance under sparsity. Overall, our model demonstrates applicability and robustness on datasets with sparse annotations per annotator.

\subsection{Qualitative Results}
\label{sec:qualitative}
We qualitatively analyze how the learned attention reflects annotators' behavior patterns by visualizing cross-attention weights from Q-Former. These weights highlight the image patches or video frames that different annotators may focus on when making predictions.

As shown in Figure~\ref{fig:S1}, on the STREET dataset (city impression classification), annotators exhibit distinct spatial focus: \emph{annotators 4, 5, 6, 7, and 8} focus more centrally on a dog in the image, while others do not. Correspondingly, these annotators also provided higher scores for the ``happiness'' dimension, suggesting that variations in focus on specific semantics may contribute to their differing judgments.

Figure~\ref{fig:A1} illustrates results on the AMER dataset (video emotion classification). Annotators differ in temporal focus: \emph{annotator 1} focuses on the final frames (45–50), while \emph{annotator 4 and 13} concentrate early frames (5–10). These patterns align with their predictions: ``sad'' for \emph{annotator 1}, ``worried'' for \emph{annotator 4 and 13}, suggesting that differences in the people or dialogues in the corresponding frame segments they focused on may underlie decision diversity.

\begin{table}
\centering
% \small
\caption{Ablation study. The average performance (accuracy) of replacing modules, removing inter-annotator correlations, and a consensus prediction comparison with and without individual annotators’ behavior pattern modeling.}
% \caption{Ablation study. The average performance of replacing some modules for annotator modeling, removing inter-annotator correlations, and a consensus prediction comparison with and without individual annotators’ behavior pattern modeling (Pre-mv vs. Post-mv), evaluated by accuracy.}
\label{tab:ablation}
\begin{tabular}{lcccccc}
\toprule
Method & S-Ha & S-He & S-Sa & S-Li & S-Or & AMER \\
\midrule
Base & 0.42 & 0.47 & 0.47 & 0.53 & 0.47 & 0.28 \\
w/ U-cls & 0.50 & 0.56 & 0.49 & 0.56 & 0.49 & 0.61 \\
% \midrule
w/o S-Attn & 0.52 & 0.54 & 0.46 & 0.53 & 0.56 & 0.73 \\
\midrule
Pre-mv & 0.58 & 0.57 & 0.56 & 0.58 & 0.60 & 0.43 \\
Post-mv & 0.62 & 0.58 & 0.61 & 0.59 & 0.62 & 0.60 \\
% \midrule
% w/ Masked & 0.48 & 0.51 & 0.48 & 0.49 & 0.47 & 0.69 \\
\midrule
Full (avg) & 0.63 & 0.64 & 0.58 & 0.62 & 0.61 & 0.84 \\
\bottomrule
\end{tabular}
\end{table}

\subsection{Ablation Study}
\label{sec:ablation_study}
We conduct an ablation study to validate our architectural design choices and assess the effect of individual behavior pattern modeling on consensus prediction (Table~\ref{tab:ablation}).
% An ablation study is conducted to validate our architectural design choices and the benefits of behavior pattern modeling for consensus prediction, as summarized in Table~\ref{tab:ablation}.

\textbf{Architecture Choices.}  
Removing Q-Former yields the Base model, and replacing individual classifiers with a unified classifier (w/ U-cls) both lead to significant performance drops, validating their effectiveness in our architecture.
% \textbf{Core Architecture Validation.} We first remove Q-Former reducing to the Base model; Second, we only use a unified classifier (w/ U-cls). Either of these decreases the overall annotator modeling performance, highlighting their effectiveness in our architecture.

\textbf{Inter-Annotator Correlation Analysis.}  
% Disabling self-attention (w/o S-Attn) in Q-Former reduces performance, especially on AMER, demonstrating the importance of capturing inter-annotator correlations as supervisory signals.
Disabling self-attention (w/o S-Attn) leads to clear performance degradation, underscoring the role of inter-annotator correlations as an implicit structural regularizer that improves generalization and robustness of individual annotator modeling.
% \textbf{Inter-annotator Correlation Analysis.} To verify the effectiveness of inter-annotator correlations as additional supervision, we remove the self-attention module (w/o S-Attn) in Q-Former. Results show performance decreases across most dimensions, demonstrating the advantage of capturing inter-annotator correlations in our architecture design.

\textbf{Individual Behavior Modeling for Consensus.}  
To investigate the role of individual behavior modeling in consensus prediction, we compare two strategies:  
(1) Pre-mv performs majority voting before modeling,\ie, applies global pooling over all Q-Former queries for a single prediction; (2) Post-mv first models each annotator individually and then aggregates their predictions.  
Post-mv consistently outperforms Pre-mv, especially on AMER.
This demonstrates that modeling individual annotators' behavior patterns preserves valuable information otherwise lost in early aggregation, potentially benefiting real-world consensus prediction or other practical applications.
% \textbf{Benefits of Behavior Pattern Modeling.} 
% To investigate whether modeling individual annotators' behavior patterns preserves valuable information for specific tasks (\eg, consensus prediction), we compare two strategies: Pre-mv, which first aggregates labels via majority voting and then performs global pooling over the entire set of Q-Former queries for consensus prediction; and Post-mv, which first models individual annotators' behavior patterns using learnable queries and independent classifiers, followed by majority voting on the resulting predictions.
% Results show that Post-mv outperforms Pre-mv overall, achieving substantial gains on AMER.
% This suggests that individual annotators' behavior patterns modeling preserves valuable information about different annotators' perspectives that would otherwise be lost in direct label aggregation, potentially benefiting consensus prediction performance.
% While this provides indirect evidence supporting the utility of annotator-specific modeling, we acknowledge that further research is needed to fully characterize its impact on various downstream applications and in scenarios beyond the evaluated datasets.
%-------------------------------------------------------------------------

\section{Conclusion}
\label{sec:conclusion}
This paper introduced a paradigm shift in multi-annotator learning from sample-wise aggregation to annotator-wise behavior modeling, and proposed a lightweight query-based architecture (\approach) to effectively model individual annotator behavior patterns. We contributed the STREET and AMER datasets with dense per-annotator labels for longitudinal annotator behavior understanding. Experiments demonstrate the superiority of our method in modeling individual annotators' behavior patterns, their utility for consensus prediction, and applicability under sparse annotations. This work presents a novel view on recognizing and understanding the challenges in the multi-annotator learning field.
%-------------------------------------------------------------------------

\begin{figure*}[t]
  \centering
  \includegraphics[width=0.9\linewidth]{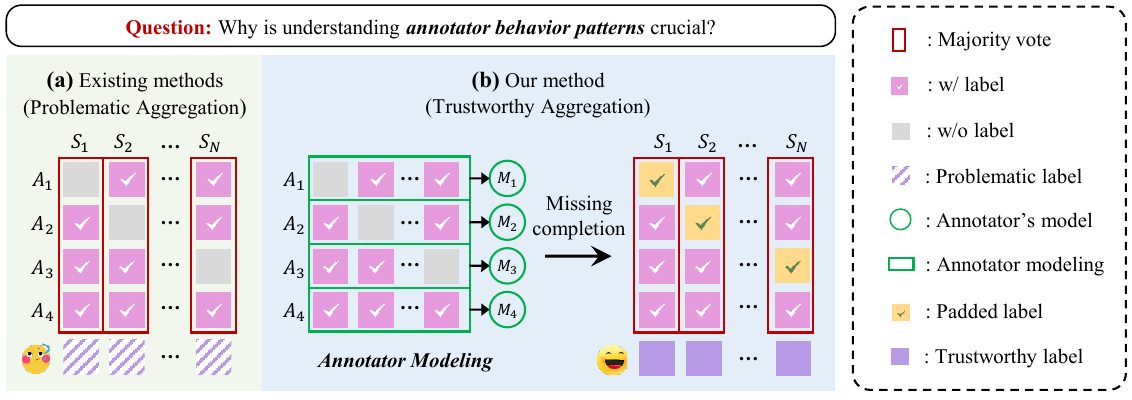}
    \caption{An illustration of paradigm shift from sample-wise aggregation to annotator-wise behavior modeling. \textbf{(a)} Existing methods overlook individual annotator information, potentially leading to suboptimal or biased consensus aggregation on real-world annotation matrices, where each annotator labels only a small subset of samples with disjoint coverage. \textbf{(b)} Our method captures each annotator's longitudinal behavior patterns across their labeled samples, improving consensus aggregation reliability by leveraging reconstructed unlabeled data.}
    \label{fig:paradigm-shift}
\end{figure*}

\begin{figure*}[t]
  \centering
  \includegraphics[width=\linewidth]{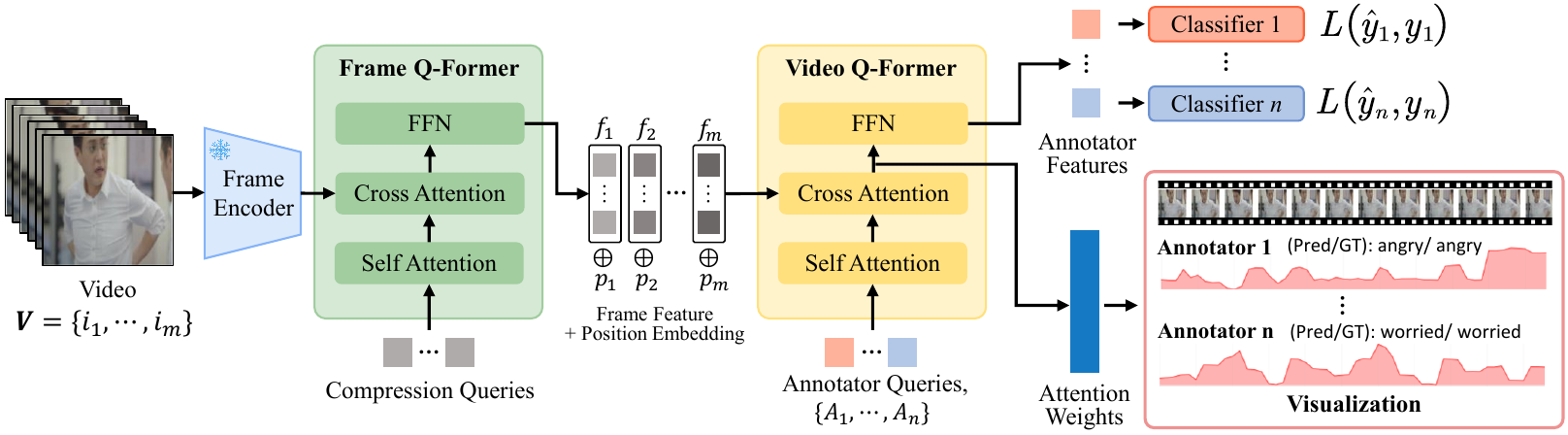}
    \caption{\approach\ architecture for video-specific pipeline. A frozen pre-trained encoder extracts frame features, which are compressed through the Frame Q-Former. With added frame position embedding, these interact with $1, \dots, n$ annotator-specific learnable queries in the Video Q-Former through cross-attention, producing annotator-specific features for classification. Different annotators’ cross-attention weights represent different annotator tendencies and provide visualization for analysis.}
    \label{fig:architecture-video}
\end{figure*}

\section{Supplementary Material}

\subsection{Overview}
First, we provide an empirical analysis, elaborating on our first contribution in the paper, \ie, practical value of our paradigm shift from sample-wise aggregation to annotator-wise modeling, further clarifying our motivation and practical contribution.
Subsequently, we illustrate the detailed video-specific architecture.
Additionally, we present supplementary experimental analysis.
The analysis of model efficiency and the discussion are given.
Finally, the discussion of limitation and future work is provided.
%-------------------------------------------------------------------------

\begin{table}[!t]
\centering
\small
\caption{Aggregation accuracy under simulated missing labels in the test set. Modeling annotators before majority voting (S-Ours-MV) outperforms direct voting (S-MV) and the best-performing baseline (S-PADL-MV), validating the benefit of annotator-wise modeling for reliable aggregation.}
\label{tab:sparse-mv}
\begin{tabular}{l@{\hspace{2.8mm}}ccccc@{\hspace{2.8mm}}c}
\toprule
Method &  S-Ha &  S-He &  S-Sa &  S-Li &  S-Or &  AMER \\
\midrule
\multicolumn{7}{c}{\emph{20\% missing rate}} \\
\midrule
S-MV & 0.97 & 0.97 & 0.96 & 0.96 & 0.96 & 0.85 \\
S-PADL-MV & 0.97 & \textbf{0.98} & \textbf{0.97} & 0.96 & \textbf{0.97} & 0.86 \\
S-Ours-MV & \textbf{0.98} & \textbf{0.98} & \textbf{0.97} & \textbf{0.98} & \textbf{0.97} & \textbf{0.89} \\
\midrule
\multicolumn{7}{c}{\emph{30\% missing rate}} \\
\midrule
S-MV & 0.94 & 0.93 & 0.92 & 0.93 & 0.92 & 0.85 \\
S-PADL-MV & 0.95 & 0.95 & \textbf{0.95} & 0.94 & 0.94 & 0.87 \\
S-Ours-MV & \textbf{0.97} & \textbf{0.96} & \textbf{0.95} & \textbf{0.96} & \textbf{0.95} & \textbf{0.92} \\
\midrule
\multicolumn{7}{c}{\emph{40\% missing rate}} \\
\midrule
S-MV & 0.92 & 0.92 & 0.89 & 0.91 & 0.90 & 0.81 \\
S-PADL-MV & 0.93 & 0.93 & 0.91 & 0.92 & 0.91 & 0.83 \\
S-Ours-MV & \textbf{0.95} & \textbf{0.95} & \textbf{0.93} & \textbf{0.94} & \textbf{0.93} & \textbf{0.88} \\
\bottomrule
\end{tabular}
\end{table}

\begin{table*}[!t]
\centering
\small
\caption{Label statistics and missing rates of the AMER dataset with 13 annotators $A_i, i \in \{1, \cdots, 5\}$. For each annotator, the number of labeled samples, the corresponding missing rate (\%), as well as the average data (Average), are provided.}
\label{tab:dataset_statistic}
\begin{tabular}{lcccccccccccccc}
\toprule
Perspective & $A_1$ & $A_2$ & $A_3$ & $A_4$ & $A_5$ & $A_6$ & $A_7$ & $A_8$ & $A_9$ & $A_{10}$ & $A_{11}$ & $A_{12}$ & $A_{13}$ & Average \\
\midrule
Sample size  & 1096 & 1031 & 1022 & 1036 & 1012 & 970 & 1064 & 1049 & 1060 & 1062 & 5187 & 5197 & 5202 & 1999.1 \\
\midrule
Missing rates (\%) & 79.0 & 80.2 & 80.4 & 80.1 & 80.6 & 81.4 & 79.6 & 79.9 & 79.6 & 79.6 & 0.4 & 0.2 & 0.1 & 69.6 \\
\bottomrule
\end{tabular}
\end{table*}

\subsection{Empirical Validation: Annotator Modeling Before Aggregation for Reliability}
\label{sec:clarification}
We present empirical validation and analysis to concretize our first contribution—the paradigm shift from sample-wise aggregation to annotator-wise behavior modeling—further clarifying our motivation and practical significance.

\textbf{Importance of Annotator Modeling for Aggregation Reliability.} As illustrated in Figure~\ref{fig:paradigm-shift}, traditional aggregation strategies operate along the sample axis without modeling longitudinal annotator behavior. In real-world sparse and non-overlapping scenarios, this leads to suboptimal or biased consensus aggregation: sample $S_1$ may be aggregated from annotators $A_2, A_3, A_4$, while sample $S_2$ from $A_1, A_3, A_4$—sometimes even entirely disjoint groups. Therefore, we hypothesize that modeling annotators individually allows for better reconstruction of their missing labels, thereby enabling more consistent and trustworthy aggregation across all samples (\eg, all $S_1$--$S_N$ aggregated from $A_1, A_2, A_3, A_4$) given well-trained annotator models.

\textbf{Validation: Direct Aggregation vs. Aggregation after Annotator-wise Modeling.}
To validate this, as shown in Table~\ref{tab:sparse-mv}, we simulate missingness on the test set by masking \{20\%, 30\%, 40\%\} of annotations and compare:
(1) \textbf{S-MV}: Direct sample-wise majority vote on remaining labels; (2) \textbf{S-Ours-MV}: Predict missing labels using annotator-specific models trained on the full training set, then aggregate all (remaining + padded) labels; Similarly, \textbf{S-PADL-MV} represents the best-performing baseline PADL~\cite{PADL} results. The results in Table~\ref{tab:sparse-mv} show that our method consistently outperforms baselines, validating that longitudinal annotator-wise modeling enhances aggregation reliability.

This section provides empirical evidence for our paradigm shift, further clarifying the motivation of this paper while depicting its practical utility value for the real world.
%-------------------------------------------------------------------------

\subsection{Video-specific Architecture}
\label{sec:architecture}
Different from the image input pipeline in the main paper, Figure~\ref{fig:architecture-video} presents the video input pipeline specifically designed for the  AMER, etc. video datasets. Specifically, given a video input $V \in \mathbb{R}^{T \times H \times W \times 3}$, a frozen pre-trained image encoder \cite{EVA-CLIP} first extracts frame features, which are then further compressed by the image Q-Former~\cite{BLIP2} using a specific number of compression queries, typically 32, to alleviate subsequent computational costs \cite{BLIP2}. The frame position embedding is then added and input to the video Q-Former. Subsequently, we assign a learnable query token in a Q-Former for modeling each annotator $A_k$ ($k = 1, \dots, n$), then all annotator-specific queries in cross-attention simultaneously interact with input features to capture their diverse tendencies and obtain the annotator-specific feature. Finally, the specific annotator features are mapped through a fully connected layer to an appropriate feature dimension and then connected to each annotator’s corresponding classifier to output classification results.

For qualitative understanding, the focus regions of the video input are based on all frames. Different annotators’ varying levels of focus on different frames indicate their behavior pattern differences: \emph{annotator 1} shows more focus on the final frames, while \emph{annotator n} focuses more on the middle frames. This dataset is for video emotion classification. Analyzing the annotation and visualization results, we see that annotators labeled emotions ``anger'' and ``worried''. The differences in the frame segments they focused on might partially contribute to the variations in their judgments.
%-------------------------------------------------------------------------

\begin{table}[htbp]
\centering
% \small
\caption{Model efficiency analysis. In this table, we report the number of parameters and the average inference time per sample. Lower values for both metrics indicate higher model efficiency.}
\label{efficiency}
\begin{tabular}{cccccc}
\toprule
Models & Parameters\thinspace(M) & Average Processing Time\thinspace(s) \\
\midrule
D-LEMA & $214.18$ & $5.64$ \\
PADL & $168.94$ & $4.59$ \\
MaDL & $201.37$ & $5.06$ \\
Ours & $\bf{106.02}$ & $\bf{4.28}$ \\
\bottomrule
\end{tabular}
\end{table}

\subsection{Model Efficiency Analysis}
\label{sec:efficiency}

This section evaluates the model efficiency from two aspects: model complexity and inference time. For model complexity, we use the number of parameters as the evaluation metric; for inference time, we compute the average processing time per sample \cite{Panoptic-tcsvt, Panoptic-wacv, Thermal-to-Color, PhD, 3DFacePolicy}. To ensure a fair comparison, Table \ref{efficiency} compares the performance of different methods in the image input pipeline (see Figure \ref{fig:architecture-image} in the main paper). Specifically, we evaluate the efficiency of different methods on one perspective of the  STREET dataset with 10 annotators. It is worth noting that different methods use different backbone networks. To ensure fairness, we standardize the backbone network to ResNet-34 for all methods. In Table \ref{efficiency}, we observe that our model has fewer parameters than the other models, while also maintaining competitive average processing time. This validates our claim in the paper that, compared to the baselines that create separate conventional models for each annotator, our architecture demonstrates significant advantages in model efficiency.

% \begin{table}[htbp]
% \centering
% \small
% \caption{Validation randomly deleting labels from STREET and AMER datasets with different proportions for scenarios of sparse continuous annotations per single annotator.}
% \label{tab:sparse}
% \begin{tabular}{llccccccccccc}
% \toprule
% Method & S-Ha & S-He & S-Sa & S-Li & S-Or & AMER \\
% \midrule
% 20\%-PADL & 0.53 & 0.52 & 0.46 & 0.51 & 0.50 & 0.80 \\
% 20\%-Ours & 0.61 & 0.60 & 0.55 & 0.58 & 0.59 & 0.88 \\
% 30\%-PADL & 0.48 & 0.47 & 0.41 & 0.46 & 0.45 & 0.76 \\
% 30\%-Ours & 0.58 & 0.57 & 0.52 & 0.55 & 0.56 & 0.85 \\
% 40\%-PADL & 0.44 & 0.42 & 0.37 & 0.43 & 0.41 & 0.73 \\
% 40\%-Ours & 0.55 & 0.53 & 0.49 & 0.52 & 0.54 & 0.82 \\
% Full-PADL & 0.58 & 0.59 & 0.52 & 0.58 & 0.55 & 0.87 \\
% Full-Ours & 0.63 & 0.64 & 0.58 & 0.62 & 0.61 & 0.92 \\
% \bottomrule
% \end{tabular}
% \end{table}

\subsection{Discussion}
\label{sec:discussion}
We supplement some extended discussions about our work here.
% First, given that existing multi-annotator datasets often have sparse continuous annotation samples per single annotator, to validate the applicability of our model in these scenarios, we conducted an experiment by randomly deleting labels from AMER and STREET with different proportions, such as 20\%, 30\%, and 40\%. As shown in Table~\ref{tab:sparse}, compared to the competing baseline PADL, we maintain superiority across different proportions, with lower performance degradation. Particularly when 40\% of labels are deleted, our average performance only decreases by 14\%, while PADL decreases by 24\%. This superiority may come from our model's ability to capture inter-annotator correlations during modeling, considering the overall hidden an additional supervisory signal among shared annotators. The experimental results indicate that our model remains effective and applicable to datasets with sparse continuous annotations from single annotators. In the future, we plan to apply and validate it on many other different real-world scenario datasets.
Our architecture design superiorly balances effectiveness and complexity, provides an accompanying visualization analysis of annotator tendencies through cross-attention weights. This has potential value and interest for understanding different annotator judgments as described in the main paper. Currently, it is introduced as an accompanying function rather than a main contribution of our paper, but in the future, we hope it can be developed into a mature interpretability solution. We plan to explore to enhance it, which may need pixel-level annotations indicating specific regions that annotators focus on during the annotation process, although this could be expensive. We could also explore quantitative evaluation ways to this further interpretability, such as feature importance ranking consistency, attention-based fidelity metrics, and human intuition consistency assessments through user studies, etc, to further enhance its value and contribution to the community.

In the ablation study of our main paper, we validated an insight that modeling multiple annotator tendencies can obtain more annotator information and demonstrated its effectiveness through experimental results based on consensus aggregation using majority voting. In future work, we plan to extend more similar experimental scenarios and validations to enhance the explanation of how multi-annotator tendency learning helps and adds benefits to specific applications \cite{uneven}.

\subsection{Additional Results}
\label{sec:results}
We provide additional experimental results about visualization analysis of annotator tendencies on both AMER and STREET datasets.

\begin{figure*}
  \centering
  \begin{subfigure}{1.0\linewidth}
    \centering
    \includegraphics[width=0.85\linewidth]{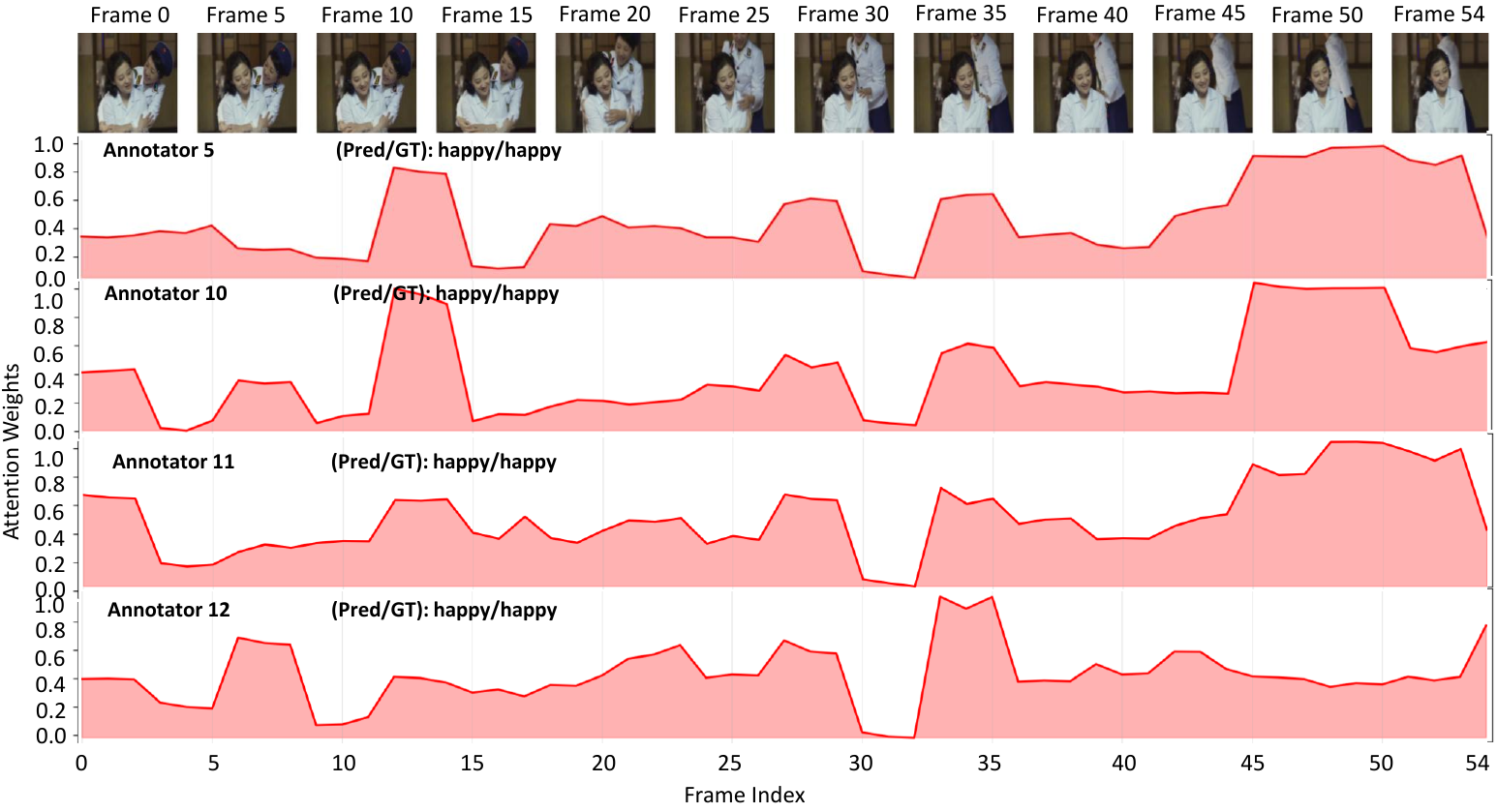}
    \caption{Sample 1.}
    \label{fig:va1}
  \end{subfigure}

  \vspace{2em}

  \begin{subfigure}{1.0\linewidth}
    \centering
    \includegraphics[width=0.85\linewidth]{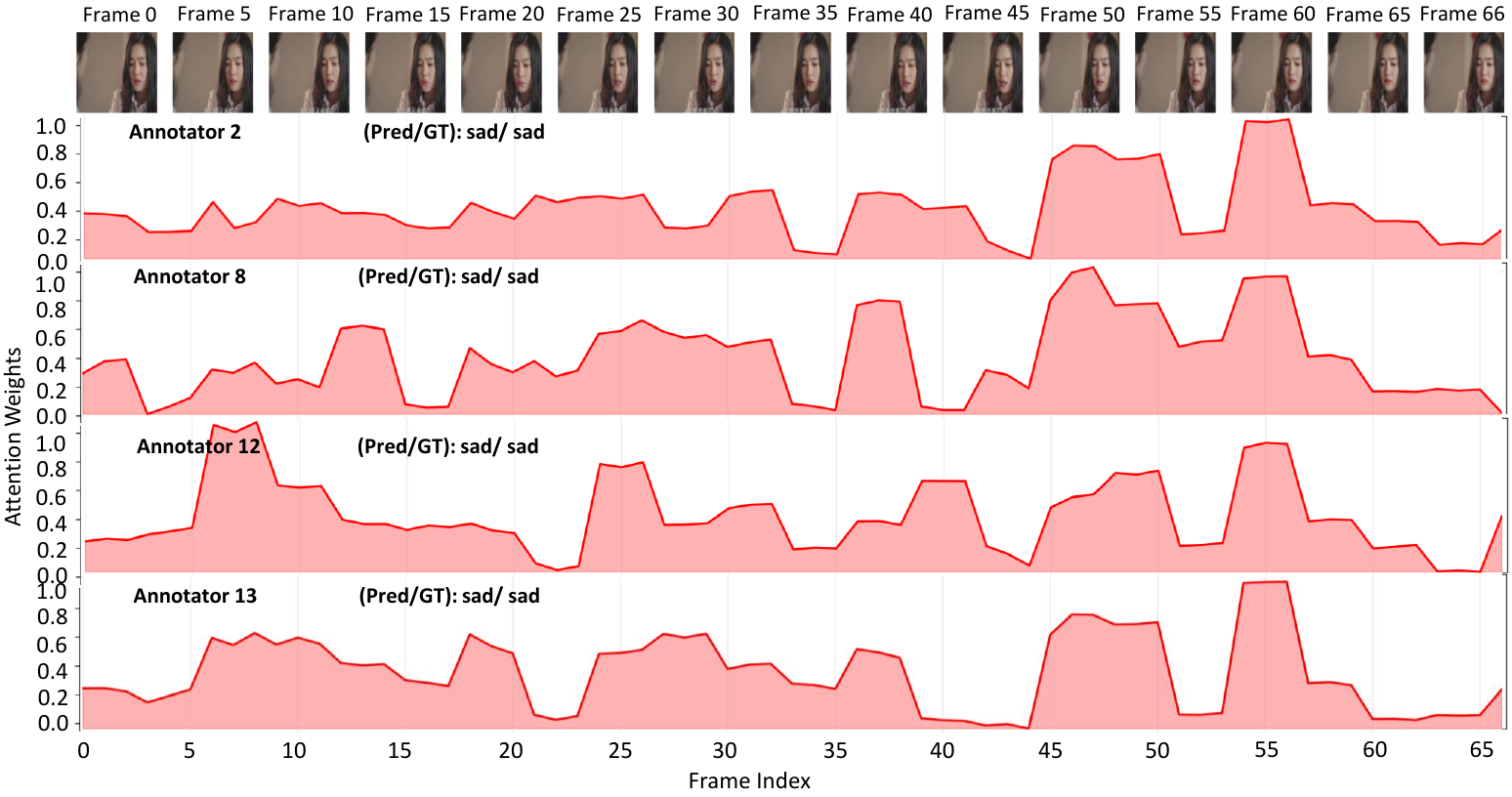}
    \caption{Sample 2.}
    \label{fig:va2}
  \end{subfigure}

  \caption{Additional sample experimental results visualize the annotator tendencies on the AMER video dataset. The tendencies of multi-annotators reveal distinct preferences.}
  \label{fig:va}
\end{figure*}

As shown in Figure~\ref{fig:va}, on the AMER dataset, annotators demonstrate different tendencies through varying focus on different frames. Sample 1 reveals the differences: \emph{annotators 5, 10, and 11} show similarly higher focus on the final frames (45-54), while \emph{annotator 12} focuses more on middle frames (30-35). They all predict the correct label ``happy'', but the different focus positions indicate that \emph{annotator 12}'s preference pattern for happiness in sample 1 differs from most annotators. Sample 2 shows: \emph{annotators 2, 8, and 13} exhibit similarly higher focus on the final frames (45-55), while \emph{annotator 12} focuses more on early frames (5-10). They all predict ``sad'' but \emph{annotator 12} demonstrates a different preference pattern.

\begin{figure*}
  \centering
  \begin{subfigure}{1.0\linewidth}
    \centering
    \includegraphics[width=0.85\linewidth]{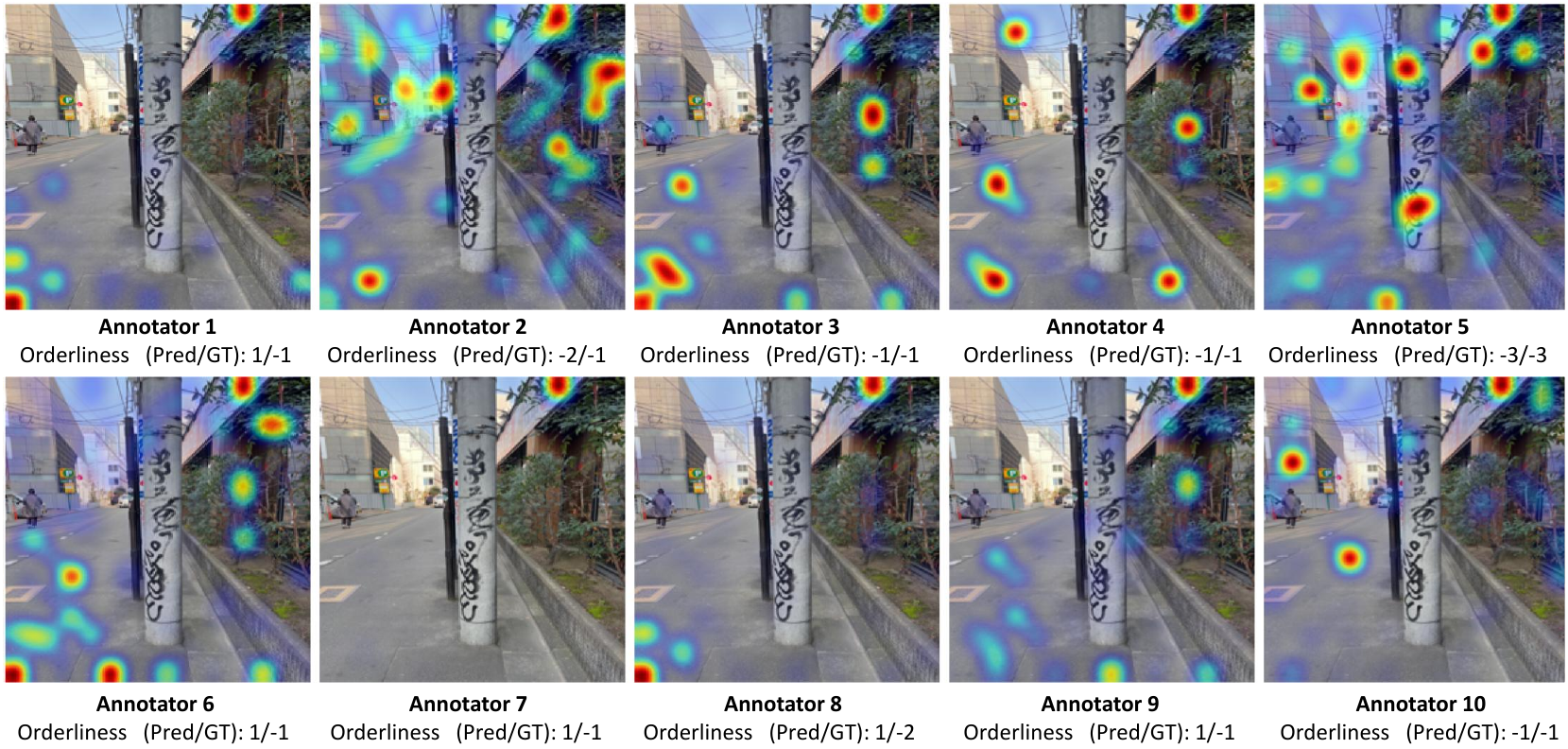}
    % \vspace{2}
    \caption{Sample 1 for orderliness perspective.}
    \label{fig:ia1}
  \end{subfigure}

  \vspace{2em}

  \begin{subfigure}{1.0\linewidth}
    \centering
    \includegraphics[width=0.85\linewidth]{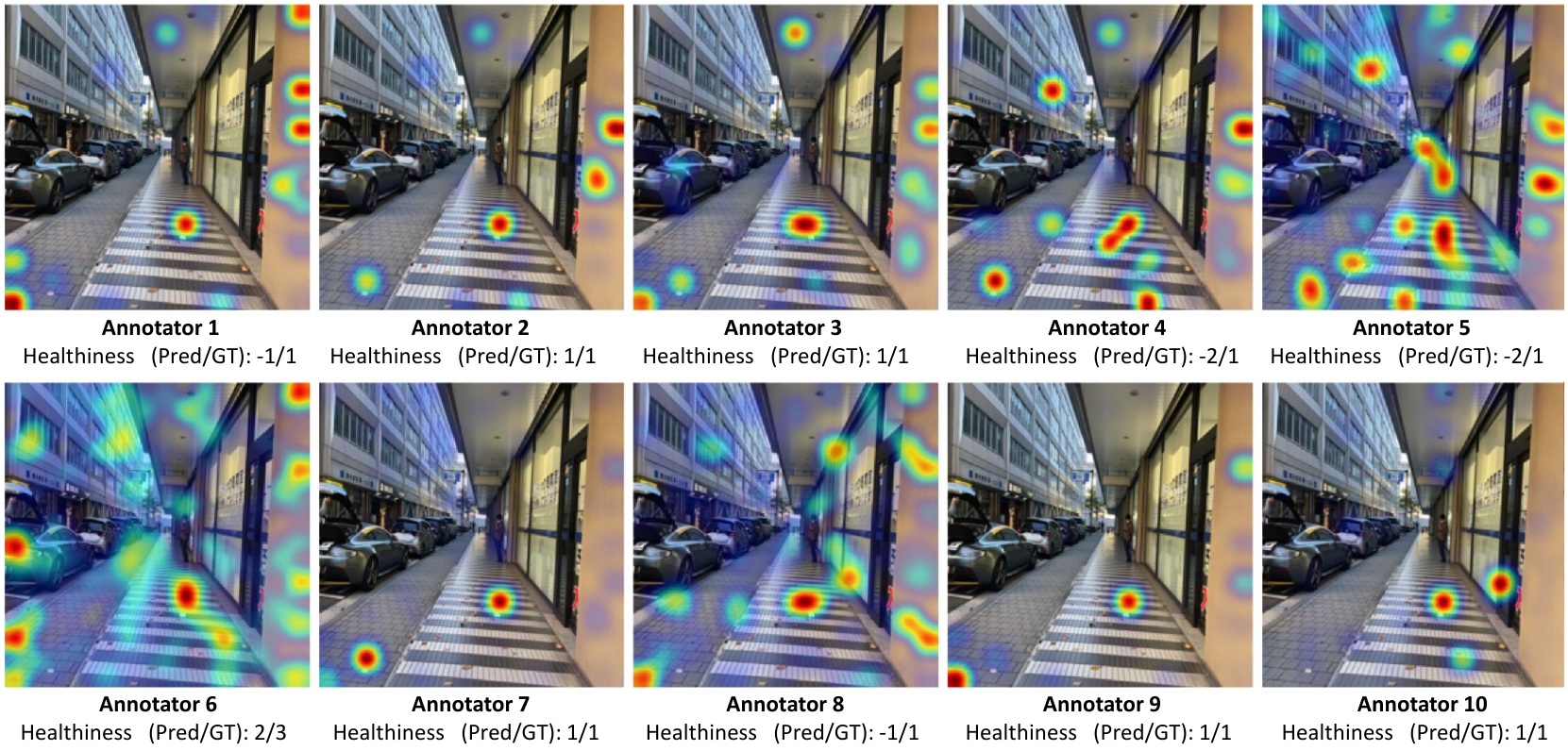}
    \caption{Sample 2 for healthiness perspective.}
    \label{fig:ia3}
  \end{subfigure}

  \caption{Additional sample experimental results visualize the annotator tendencies on the STREET image dataset. The tendencies of 10 annotators reveal distinct preferences.}
  \label{fig:ia}
\end{figure*}

As shown in Figure~\ref{fig:ia}, on the STREET dataset, annotators exhibit different tendencies through varying focus on different semantic elements within the same input image. In sample 1 from the orderliness perspective, the focus regions show differences: \emph{annotator 5} focuses more intensively on graffiti in the image compared with other annotators. The results show that prediction and ground truth are highly consistent in the emotion reflected in the semantics, corresponding to this annotator's expected lowest score. In sample 2 of the healthiness perspective, most annotators focus on uncleaned garbage and leaves, and only \emph{annotator 6} shows more focus on the surrounding environment, like cars and buildings, which might influence the result of not giving a low score.

\bibliographystyle{ACM-Reference-Format}
\bibliography{sample-base}

\end{document}